\documentstyle[psfig]{elsart}

\journal{Physica A}

\begin{document}

\begin{frontmatter}
\title{Quantum Frenkel-Kontorova Model}

\author{Bambi Hu}
\address{Department of Physics and Centre for Nonlinear Studies, 
Hong Kong Baptist University, Hong Kong, China\\
Department of Physics, University of 
Houston, Houston, TX 77204-5506, USA} 

\author{Baowen Li}
\address{Department of Physics and the Centre for Nonlinear Studies, 
Hong Kong Baptist University, Hong Kong, China \\
Department of Physics, National
University of Singapore, 119260 Singapore}

\begin{abstract}

This paper gives a review of our recent work on the quantum
Frenkel-Kontorova model. Using the squeezed state theory and the quantum
Monte Carlo method, we have studied
the effects of quantum fluctuations on the Aubry transition and the
behavior of the ground state wave function. We found that quantum
fluctuations renormalize the sinusoidal standard map to a
sawtooth map. Although quantum fluctuations have smeared the Aubry
transition, the remnants of this transition are still discernible. The
ground state wave function also changes from an extended state to a
localized state. The squeezed state results agree very well with those
from the Monte Carlo and mean field studies.

\end{abstract}


\end{frontmatter}

\section{Introduction}


Many physical systems such as charge density waves, magnetic spirals,
Josephson junctions and adsorbed monolayers exhibit a competition between
two or more length scales. To understand such modulated structures, a
simple model, known as the Frenkel-Kontorova (FK) model \cite{FK38}, was
introduced
more than half a century ago. The FK model describes a chain of atoms
connected by springs in the presence of an external sinusoidal potential.
It has two length scales: the natural length of the spring and the period
of the external potential. Due to the competition between these
two length scales, the FK model exhibits a wealth of interesting and
complex phenomena. Extensive studies of the FK model have been made since
its
introduction. In most of the earlier studies, the FK model was treated in
the
continuum approximation. Although the continuum approximation, which
reproduces the sine-Gordon equation and its soliton solutions, gives some
very useful information, it misses many essential features. A major
advance in the study of the FK model was made by Aubry \cite{Aubry}.  
Aubry reverted to
the discrete version of the FK model and made essential use of the seminal
Kolmogorov-Arnol'd-Moser (KAM) theorem. In analogy to the breakup of a KAM
torus, Aubry discovered the ``transition by breaking of analyticity.'' The
Aubry transition is similar to a  phase transition, thus all the apparatus
invented in the study of second-order phase transitions can be borrowed to
study this transition. A fairly complete picture of the FK model was
thus obtained. 

So far most of the studies made on the FK model have been confined to
the classical model. Very little study has been done on the quantum FK
model. Although the FK model looks simple, it's nevertheless a
non-integrable interacting
many-body system. There aren't that many methods
at
our disposal to study the quantum behavior of such a model. Moreover,
besides theoretical interest, there are also practical considerations
that demand a deeper understanding of the quantum FK model. For
example, the
study of friction, or tribology \cite{tribology}, is a very important
subject both in terms
of fundamental science and applications. And indeed, the FK model has been
used to study tribology. However, as technology is moving into the
nano-regime, nanotribology \cite{Klafter} is also becoming a very
important subject of
study. In the nano-regime, it's more likely than not that quantum
mechanics
will play an important role. This makes an understanding of the quantum FK
model not only desirable but very likely imperative. This is the
motivation for our interest in the quantum FK model.

The first study of the quantum FK model was made by Borgonovi, Guarneri
and Shepelyansky \cite{BGS89}. They mainly used a numerical approach.
Later,
Berman, Bulgakov and Campbell \cite{BBC94} studied the same problem using
a mean-field approach. In our study we will use a hybrid analytical and
numerical approach. The analytical method is based on the squeezed state
theory, and the numerical method is based on the quantum Monte Carlo (QMC)
algorithm.

This paper gives a review of our recent work on the quantum FK model
\cite{HLZ98,HL99}. 
It is organized as follows. In Sec. 2, we will define the quantum
Frenkel-Kontorova model and use the Monte-Carlo method to study it. In
Sec. 3 we will formulate the squeezed state
approach. In Sec. 4 we will study the quantum effects on the Aubry
transition. In Sec. 5 some concluding remarks will be given.

\section{Quantum Frenkel-Kontorova model}

The Hamiltonian of the quantum Frenkel-Kontorova model is described by:
\begin{equation}
\label{eq:QMHam1}
\hat{\cal H} = \sum_i \left[\frac{\hat{p}_i^2}{2m} + 
\frac{\gamma}{2}(\hat{x}_{i+1} - \hat{x}_{i} - a)^2 -  
V\cos(q_0\hat{x}_i)\right] . 
\end{equation}
Here $m$ is the mass of the particles, $\gamma$ the elastic constant of
the spring, $2\pi/q_0$ the period of the external potential, $V$ the
magnitude of the external potential, and $a$ the equilibrium distance
between two nearest neighbor particles. For convenience, we shall rescale
the variables so as to obtain a dimensionless Hamiltonian
\begin{equation}
\label{eq:QMHam} 
 \hat{H} = \sum_i \left[\frac{\hat{P}_{i}^{2}}{2} +
\frac{1}{2}(\hat{X}_{i+1} - \hat{X}_{i} - \mu)^2 -
K\cos\hat{X}_i\right], 
\end{equation} 
where  $\hat{X}_i = q_0 \hat{x}_i$, $\hat{P}_i=
(\frac{q_0}{\sqrt{m\gamma}})\hat{p}_i$, $\mu = q_0 a$, and
$K=Vq_0^2/\gamma$.
We define an ``effective Planck's constant'':
\begin{equation}
\label{eq:Nhbar}
\tilde{\hbar} =\left(\frac{q_0^2}{\sqrt{m\gamma}}\right)\hbar.
\end{equation} 
This effective Planck's constant 
is the ratio of the
natural quantum energy scale $\hbar\omega_0$ to the natural classical
energy scale $\gamma/q^2_0$, and $\omega^2_0=\gamma/m$. 
This quantity provides a measure of the importance of quantum effects.
In 
particular, quantum effects become crucial at zero temperature when 
thermal fluctuations vanish.

In the classical FK model, the positions of the particles 
in the ground state, denoted by $X_i^e$, are determined by the following 
equation:
\begin{equation}
\label{eq:GSEQ}
X^e_{i+1} - 2X^e_i + X^e_{i-1} = K\sin X^e_i.
\end{equation}
This equation can be cast into the form of the standard map by defining 
$I_{i+1} = X^e_{i+1}- X^e_i$:
\begin{equation}\begin{array}l
\label{eq:STDMP}
I_{i+1} = I_i + K \sin X^e_i , \\
X^e_{i+1} = I_{i+1} + X^e_i .
\end{array}
\end{equation}
To describe the transition from the unpinned phase in the subcritical
regime
to the pinned phase in the supercritical regime, the hull function is
introduced.
It is defined by
\begin{equation}
\label{eq:CLhull}
X^e_i = f(X_i).
\end{equation}
Here $X_i$ is the unperturbed ($K=0$) particle position. 
Another useful function is called the $g$-function,  defined by
\begin{equation}
\label{eq:CLGfunc}
g= \frac{1}{K} 
(X^e_{i+1} - 2 X^e_i + X^e_{i-1})
\end{equation}

\begin{figure}
\centerline{\psfig{file=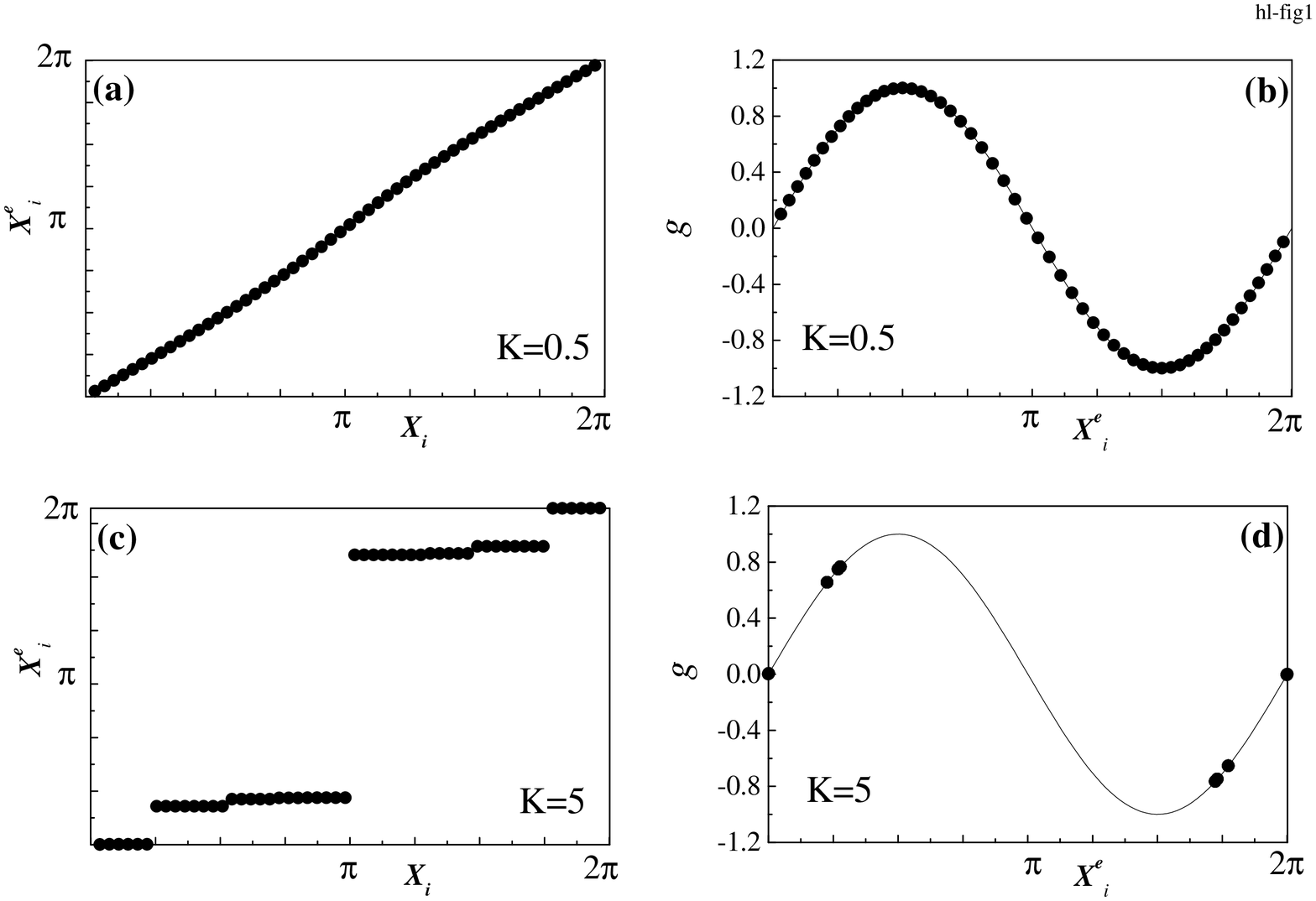,width=5.5in}}
\caption{
The classical hull function and $g$-function
in the 
subcritical regime ($K=0.5$) and the supercritical regime ($K=5$). In (b)
and (d), the solid line is $\sin X^e$.
}
\label{hullcl}
\end{figure}

For $K<K_c$,
the ground 
state of the classical FK model corresponds to
an invariant curve in the standard map (\ref{eq:STDMP}).
The hull function is a monotonic analytical function [Fig. 
\ref{hullcl}(a)], and the $g$-function is $g(X) = \sin X$ [Fig. 
\ref{hullcl}(b)]. 
At $K=K_c$, the last invariant 
curve in the standard map breaks up and becomes a Cantorus. The
hull functions becomes a monotonic function with a countable number of 
discontinuities [Fig. \ref{hullcl}(c)], and the $g$-function becomes a 
Cantor subset of the function $\sin X$ [Fig. \ref{hullcl}(d)].

By using the quantum Monte Carlo method, we can calculate 
the ground state averages of the particle positions $\bar{X}_i$ and 
plot the following three functions so as to compare with their classical
counterparts: (1) The quantum hull 
function, i.e., $\bar{X}_i$ as a function of the unperturbed 
positions $X_i$; (2) the quantum $g$-function which is defined as
\begin{equation}
\label{eq:Gfunc}
g_q = \frac{1}{K} (\bar{X}_{i+1} - 2\bar{X}_i + \bar{X}_{i-1} ) ;
\end{equation}
and  (3) ($\bar{X}_i, \bar{I}_i$) in the phase
space of the standard map. 

The quantum effect does not destroy the functional dependence of $g_q$ on
$\bar{X}_i$, instead it renormalizes it to a sawtooth
map [Fig. \ref{hullqmc}]. This phenomenon was first
observed by 
Borgonovi {\it et al.} \cite{BGS89} and was
interpreted as a
consequence of quantum
tunneling between the
edges of the gap in the hull function.


\section{Squeezed state approach}

In this section, we will study the effect of quantum fluctuations on
the FK model by using the squeezed state approach 
\cite{HLZ98}.  

The squeezed state was first used in quantum optics and has
proven to be very useful in many other fields
\cite{ZF905,Tsue91,Tsue92,PS97,HLLZ98}.
The 
squeezed state approach 
starts from the time-dependent variational principle
\begin{equation}
\delta \int dt \langle\Psi(t)|i\hbar\frac{\partial}{\partial t}
-\hat{H}|\Psi(t)\rangle = 0.
\label{TDVP}
\end{equation}
Variation with respect to (w.r.t.)  $\langle\Psi(t)|$ 
gives rise to
the Schr\"odinger equation. The exact solution may be approximated
by restricting  
the choice of states to a subspace of the full Hilbert space.
In the squeezed state approach, the squeezed state is chosen as
$|\Psi(t)\rangle$. 

The squeezed state 
$|\Phi(\alpha,\beta)\rangle$ is 
defined by
\begin{equation}
|\Phi(\alpha,\beta)\rangle = e^{\hat{S}(\alpha)} e^{\hat{T}(\beta)}|0\rangle,
\label{SS}
\end{equation}
where
\begin{equation}
\hat{S}(\alpha) = \sum_i(\alpha_i \hat{a}^\dagger_i - \alpha^*_i 
\hat{a}_i)
\label{ST1}
\end{equation}
is the Weyl's displacement operator, and 
\begin{equation}
\hat{T}(\beta) = \frac{1}{2}\sum_{ij}(\hat{a}^\dagger_i\beta_{ij} 
\hat{a}^\dagger_j - \hat{a}_i \beta^\dagger_{ij} \hat{a}_j).
\label{ST2}
\end{equation}
is the squeezed state operator. 
$\hat{S}^\dagger(\alpha) 
= - \hat{S}(\alpha), \hat{T}^\dagger(\beta) = - \hat{T}(\beta)$.
For brevitiy, we will use the 
abbreviation $|\Phi\rangle \equiv |\Phi(\alpha,\beta)\rangle$.
If $\beta =0 $, the squeezed state 
is reduced to the coherent state. However, as we shall see 
later, the coherent state is not adequate to study quantum
fluctuations.

The position and momentum operators for the $i$th particle are written as
\begin{equation}\begin{array}l
\hat{X}_i = \sqrt{\frac{\tilde{\hbar}}{2}} ( \hat{a}^\dagger_{i} +
\hat{a}_i),\\
\hat{P}_i = i \sqrt{\frac{\tilde{\hbar}}{2}} ( \hat{a}^\dagger_{i} -
\hat{a}_i) .
\end{array}
\label{xpoprator}
\end{equation}
Here, $\hat{a}^\dagger_i$ and $\hat{a}_i$ are the creation and
annihilation 
operators which satisfy the canonical commutation relations: 
$[\hat{a}_i,\hat{a}_j^\dagger]=\delta_{ij}$, $[\hat{a}_i,\hat{a}_j]=0$ and 
$[\hat{a}_i^\dagger,\hat{a}_j^\dagger]=0$.

Using $|\Phi\rangle$ as a trial wave function, we can  
find the expectation values of the position 
and momentum operators of the $i$th particle \cite{ZF905}:
\begin{equation}\begin{array}l
\bar{X}_i  \equiv  \langle\Phi|\hat{X}_i|\Phi\rangle = 
\sqrt{\frac{\tilde{\hbar}}{2}} (\alpha^*_i + \alpha_i),\\
\bar{P_i}  \equiv  \langle\Phi|\hat{P}_i|\Phi\rangle= -i 
\sqrt{\frac{\tilde{\hbar}}{2}} (\alpha^*_i - \alpha_i).
\end{array}
\label{expect}
\end{equation}
Fluctuations in the position and the momentum are given by
\begin{eqnarray}
\Delta X^2_i  &\equiv  &\langle\Phi|(\hat{X}_i - \bar{X}_i)^2|\Phi\rangle = 
\tilde{\hbar} G_{ii},\nonumber\\
\Delta P^2_i  &\equiv & \langle\Phi|(\hat{P}_i - 
\bar{P}_i)^2|\Phi\rangle = 
\tilde{\hbar} \left(\frac{G^{-1}_{ii}}{4} + 
4\sum_{l,k}\Pi_{il}G_{lk}\Pi_{ki}\right).
\label{fluct}
\end{eqnarray}
The fluctuation covariance 
between the $i$th particle and the $j$th particle is
\begin{equation}
\Delta X_i \Delta X_j \equiv \langle\Phi|(\hat{X}_i - \bar{X}_i)(\hat{X}_j - 
\bar{X}_j)|\Phi\rangle = \tilde{\hbar} G_{ij},
\label{covar1}
\end{equation}
where
\begin{equation}\begin{array}l
G_{ij}  =  \frac{1}{2} (\cosh^2\sqrt{\beta\beta^\dagger} +
\sinh^2\sqrt{\beta\beta^\dagger})_{ij}
+ \frac{1}{2}(M\beta+ \beta^\dagger M)_{ij},\\
\Pi_{ij} = \frac{i}{4}G^{-1}_{ij} (M\beta-\beta^\dagger M)_{ij},
\end{array}
\label{GPI}
\end{equation}
and,
\begin{equation}
M= 
\frac{\sinh\sqrt{\beta\beta^\dagger}\cosh\sqrt{\beta\beta^\dagger}}
{\sqrt{\beta\beta^\dagger}}.
\label{Matrix}
\end{equation}
Since $\beta$ is a symmetric matrix, $G_{ij} = G_{ji}$ and  
$\Pi_{ij} = \Pi_{ji}$. 
Using the following relation
\begin{equation}
\langle\Phi|\cos\hat{X}_i|\Phi\rangle 
=\exp\left(-\frac{\tilde{\hbar}}{2}G_{ii}\right)  \cos\bar{X}_i,
\label{cos}
\end{equation}
we obtain the expectation value of the Hamiltonian, $\bar{H}
\equiv  \langle\Phi|\hat{H}|\Phi\rangle$,
 
\begin{eqnarray}
\bar{H} 
& = & \sum_i
\frac{1}{2}\left(\bar{P}_i^2 + \tilde{\hbar}(\frac{G^{-1}_{ii}}{4} + 
4 \sum_{l,k}\Pi_{il}G_{lk}\Pi_{ki})\right)
+  \sum_i \frac{1}{2} (\bar{X}_{i+1} -\bar{X}_i - \mu)^2\nonumber\\
& + & \sum_i \frac{\tilde{\hbar}}{2} \left(
G_{ii} + G_{i+1 i+1} - 2 G_{i+1 i} 
\right)
 -  \sum_i K\exp\left(-\frac{\tilde{\hbar}}{2}G_{ii}\right)  
\cos\bar{X}_i .
\label{ExpcHam}
\end{eqnarray}

The variables $\bar{X}_i$ and $\bar{P}_i$, and $ 
G_{ij}$ and $\Pi_{ij}$ form explicitly canonical conjugates \cite{Tsue91}. 
To find the 
ground state of the quantum FK model, we will use a variational approach
in which these four variables are treated as independent
variables.  Variation with respect to 
$\bar{P}_i$  and 
$\bar{X}_i$ yields, respectively,

\begin{equation}
\begin{array}l
\bar{P}_i = 0,\\
\bar{X}_{i+1} - 2
\bar{X}_{i} +
\bar{X}_{i-1}  =  K_i \sin\bar{X}_i ,
\end{array}
\label{position}
\end{equation}
where 
\begin{equation}
K_i = K \exp\left(-\frac{\tilde{\hbar}}{2}G_{ii} \right) . 
\label{Ksubi}
\end{equation}
This equation determines the expectation value of the particle's
position. 
Unlike its classical counterpart ($\tilde{\hbar} =0$, 
$K_i = K$), this equation is 
coupled to the quantum fluctuation by $\tilde{\hbar}G_{ii}$.
Variation w.r.t. $\Pi_{ij}$ leads to
\begin{equation}
\sum_k G_{ik}\Pi_{kj} =0. 
\label{GPIsum}
\end{equation}
To obtain 
the equation 
for $G_{ij}$, we first take variation w.r.t. $G_{ik}$ and note that 
\begin{equation}
\frac{\delta G_{ij}}{\delta G_{kl}} = \delta_{ik} 
\delta_{jl} .
\label{Delta}
\end{equation}
We
then 
multiply both sides of the equation by $G_{kj}$ and sum over 
$k$. We then get the closed set of equations for the covariance $G_{ij}$:
\begin{equation}
(GF)_{ij} =  G_{i-1 j} + G_{i+1 j},
\label{covar}
\end{equation}
where
\begin{equation}
F_{ij}  =  \left( 1 
+ \frac{K_i}{2}\cos\bar{X}_i\right)  \delta_{ij} - 
\frac{(G^{-2})_{ij}}{8}.
\label{F}
\end{equation}
This is a set of equations determining the quantum fluctuations of 
the particles. $G$ is an $N\times N$ symmetric matrix which provides 
information about fluctuations. Its diagonal elements give
the variance of
each particle  while its off-diagonal elements give the covariance
between particles. From this information we can calculate the correlation
function of
the quantum fluctuations. These equations are coupled to the
expectation values $\bar{X}_i$. 

Up to this point, we have obtained $\frac{1}{2} N\times(N+1) + N$
equations for 
all the variables. These equations provide some qualitative information
about the system.
If we define $\bar{I}_{i+1} = \bar{X}_{i+1} -
\bar{X}_i$,
Eq. (\ref{position}) can be cast into the form of a standard map,
\begin{equation}\begin{array}l
\bar{I}_{i+1} = \bar{I}_i + K_i \sin\bar{X}_i,\\
\bar{X}_{i+1} = \bar{I}_{i+1} + \bar{X}_i.
\end{array}
\label{Xmap}
\end{equation}
Similarly, if we define $Q_{i+1j} = G_{i+1j} -G_{ij}$, 
Eq.(\ref{covar}) can also be cast into the form of a map:
\begin{equation}\begin{array}l
Q_{i+1j} = Q_{ij} + [G(F - 2)]_{ij},\\
G_{i+1j} = G_{ij} + Q_{i+1j}.
\end{array}
\label{Gmap}
\end{equation}
The difference between the classical 
case ($\tilde{\hbar}=0$) 
and the quantum case is readily seen in Eq.(\ref{Xmap}) . In
the classical 
case the coupling constant
does not change from potential to potential; however,
in the quantum case,
the effective coupling constant
changes from particle to 
particle. Because  $G_{ii} > 0$ for any $\tilde{\hbar} \ne 0$, $K_i<K$,
which implies that the quantum fluctuation will reduce the coupling
strength. Another important difference is that in 
the classical case the positions of the particles in the ground state are 
determined by the standard map whereas in the quantum case 
they are determined by $(N+1)$-coupled two-dimensional maps. This 
makes 
the quantum FK model extremely difficult to study analytically.

\begin{figure}
\centerline{\psfig{file=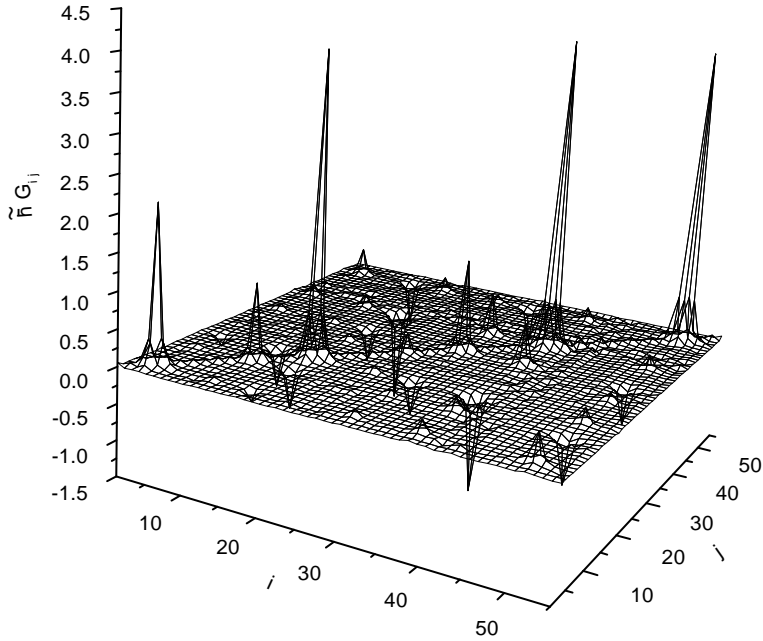,width=5.5in}}
\vspace{-.5cm}
\caption
{
$\tilde{\hbar}G_{ij}$ as a function of $i$ and $j$. The data are
obtained from QMC with $\tilde{\hbar}=0.2$,
$K=5$. 
}
\label{hgij}
\end{figure}

One notices that when $\beta =0$, $G_{ij} 
=1/2$,
which is the result of the coherent state theory. This is not the case
in the quantum FK model. In Fig. \ref{hgij} we 
plot $\tilde{\hbar}G_{ij}$ as a function of $i$ and $j$. The data are
obtained from QMC.
It is obvious that $G_{ij} \ne 1/2$. Thus it seems that the coherent
state is not adequate for the study of the quantum FK model. 

We now make some comparisons with the  QMC results. As mentioned before,
to find the solution of these two sets of
equations 
[Eqs.(\ref{position}) and (\ref{covar})] is
equivalent to finding a periodic orbit in the $2(N+1)$-dimensional map 
[Eqs.(\ref{Xmap}) and (\ref{Gmap})]. This is a very difficult problem to
solve even numerically. 
Nevertheless, we will try to solve Eq.(\ref{position}) self-consistently 
to see whether this 
equation can indeed give rise to the sawtooth map \cite{BGS89}.

\begin{figure}
\centerline{\psfig{file=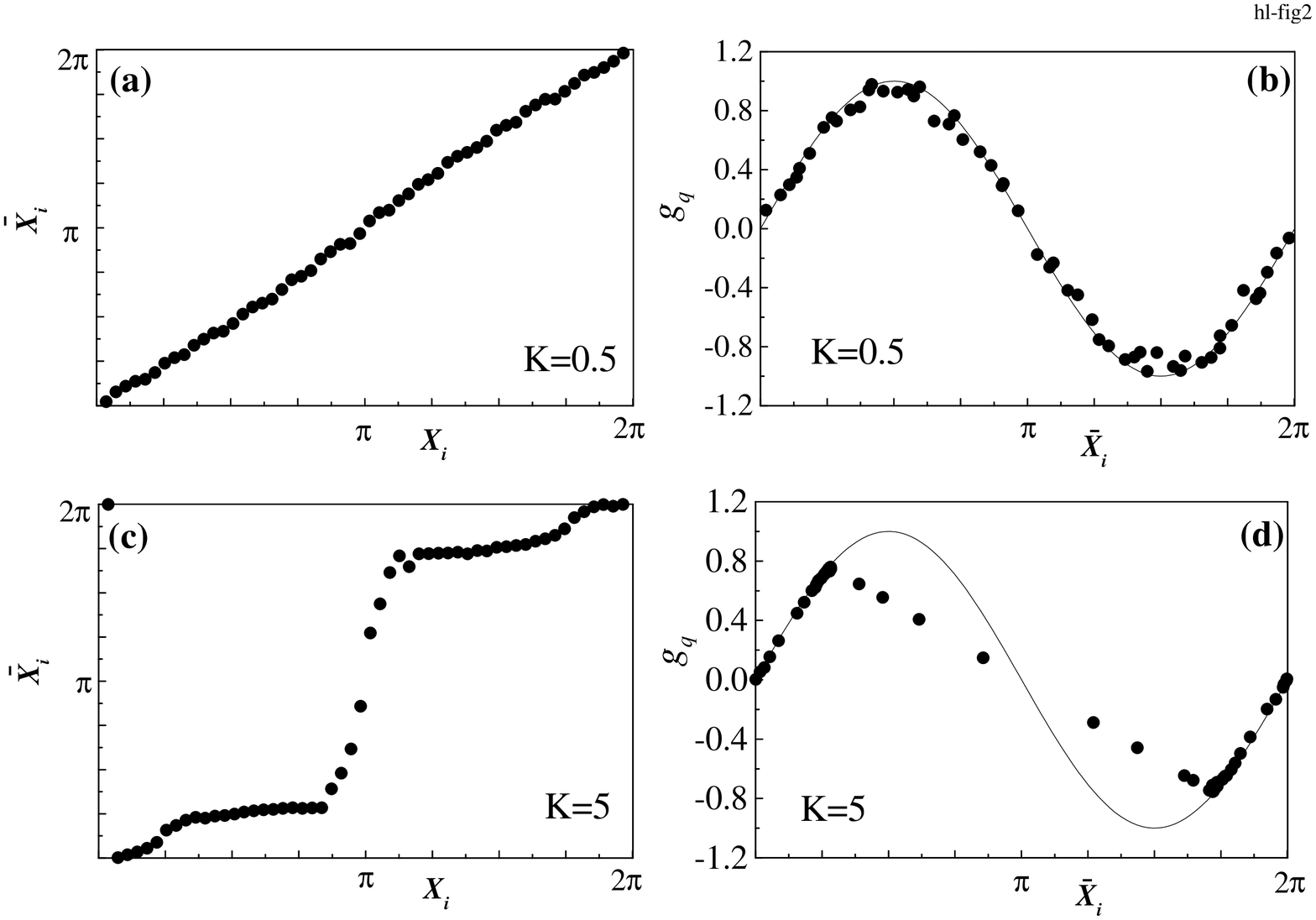,width=5.5in}}
\vspace{-0.5cm}
\caption{
The quantum hull function and $g$-function from QMC, 
$\tilde{\hbar}=0.2$. The solid line in (b) and (d) is  $\sin X$.
}
\label{hullqmc}
\end{figure}

\begin{figure}
\centerline{\psfig{file=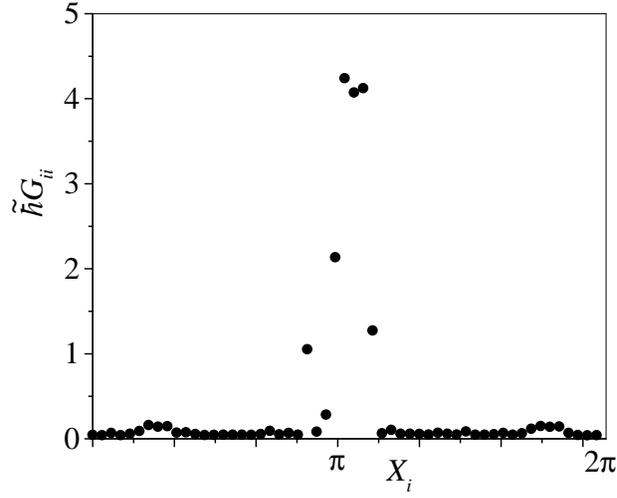,width=4.in}}
\vspace{-6.5cm}
\caption{
$\Delta X_i^2 = \langle(X_i-\bar{X}_i)^2\rangle =
\tilde{\hbar}G_{ii}$,
calculated from QMC. The 
parameters are the same as in Fig. \ref{hullqmc}. 
}
\label{qmcgii}
\end{figure}

Using QMC, we obtained the expectation value of
the particle's position $\bar{X}_i$, from which we can construct the 
quantum hull function $f_q$, as shown in Fig.
\ref{hullqmc}(a) for $K=0.5$ and (c) for $K=5$, respectively. The quantum
$g_q$-function is shown in Fig.
\ref{hullqmc}(b) and (d) for $K=0.5$ and $K=5$, respectively.  The quantum
fluctuation $\tilde{\hbar}G_{ii}$ calculated from QMC is shown in Fig.
\ref{qmcgii} for the case $K=5$.

As usual, we use $Q$ particles and $P$ external potentials with period 
$2\pi$, and the periodic boundary condition is imposed \cite{Linote}:
$\bar{X}_{Q+i} = \bar{X}_{i} + 2\pi P $. The results shown in the figure
are for $P/Q=34/55$.

\begin{figure}
\centerline{\psfig{file=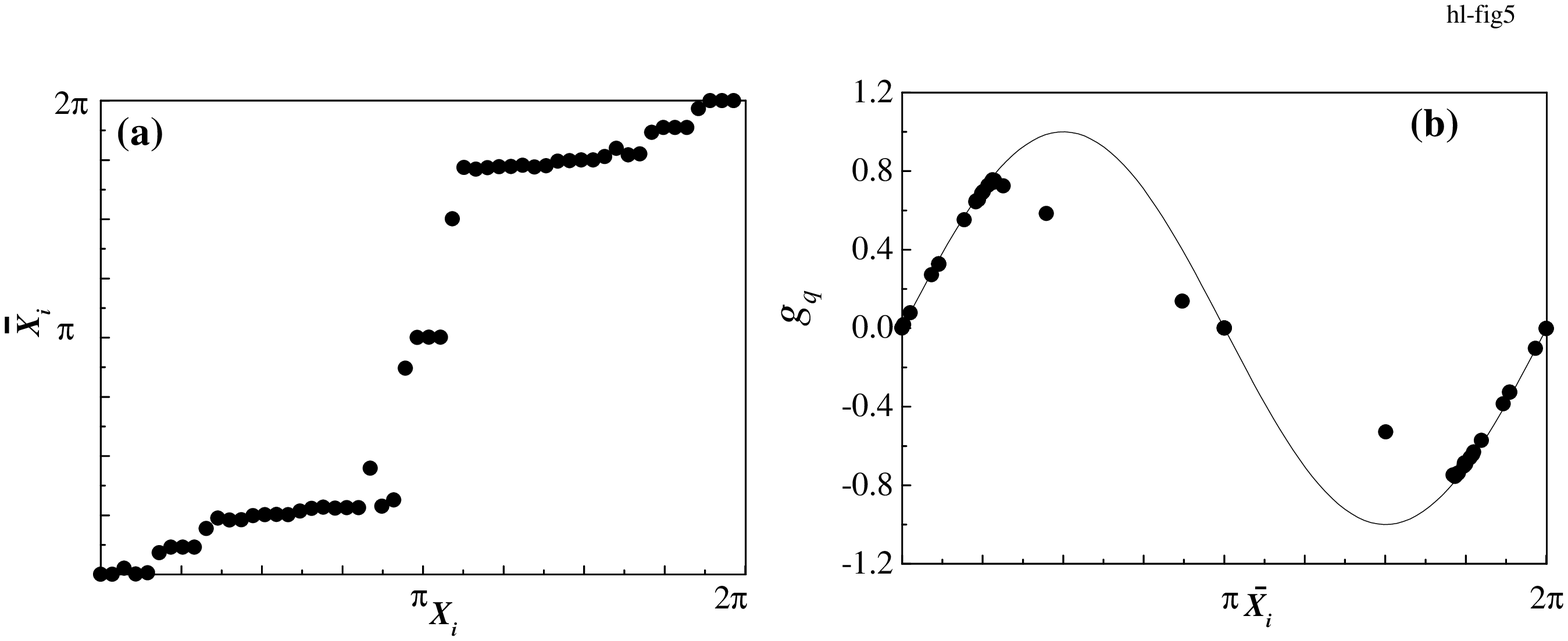,width=5.5in}}
\vspace{-3.5cm}
\caption
{
The quantum hull function and $g$-function calculated from the squeezed
state theory by
substituting $\tilde{\hbar}G_{ii}$ shown in Fig. \ref{qmcgii}
into Eq.(\ref{position}).
}
\label{hullscs}
\end{figure}

To compare the squeezed state results with those from QMC, we substitute 
$G_{ii}$ calculated from QMC into Eq.(\ref{position}). We then 
compute the expectation value of the particles' positions by using 
Aubry's gradient 
method. We then construct the quantum hull function $f_q$ and the quantum 
$g_q$-function shown in Fig. \ref{hullscs}.
The results  agree
surprisingly well with those from QMC.  
The most striking feature
is the sawtooth shape of the $g_q$-function.
This phenomenon was 
first observed by Borgonovi {\it et al.} \cite{BGS89} in their QMC 
computation and 
has been explained as a tunneling effect. Later, Berman {\it et al.} 
\cite{BBC94} obtained this result by using a mean field 
theory. In the framework of the squeezed state theory, the
renormalization of the standard map to a  
sawtooth map is a consequence of quantum fluctuations. 
Our results show that the squeezed state approach 
indeed captures the effects of quantum fluctuations. 

\section{Quantum effects on the Aubry transition}

In this section we will study the effects of quantum fluctuations on
the Aubry transition. 
We will first study the behavior of the wave function of an incommensurate
ground state. We will then introduce various indicators that might be used
to characterize the quantum Aubry transition.

\subsection{Ground state wave function of the quantum Frenkel-Kontorova
model}

We will use the QMC \cite{QMC} to study the ground state of
Hamiltonian (\ref{eq:QMHam}).  As in the classical case,
we will concentrate our attention on the incommensurate state
characterized by the golden mean winding number $\omega_G=(\sqrt{5}-1)/2$.
In the
classical case the winding number is defined as the average separation of
atoms per period, i.e.,  
$\omega=\lim_{N\rightarrow\infty}\frac{X_N-X_0}{2N\pi}$. If $\omega$ is a
rational number, the ground state is commensurate; and if it is an
irrational number, the ground state is incommensurate.  In the
quantum case, $\bar{X}_i $ is defined as the expectation
value of the position of the $i$th particle.  As usual, we use the method
of continued
fraction expansion and approximate $\omega_G$ by its rational convergents
$\omega_n = F_n/F_{n+1}$, where $F_n$ is the $n$th Fibonacci number
defined by the recursion relation $F_{n+1} = F_{n} + F_{n-1}$ with
$F_0=0, F_1=1$. In our computation, we
choose $F_{n+1}$ particles embedded in $F_n$ external
potentials. 

\begin{figure}
\centerline{\psfig{file=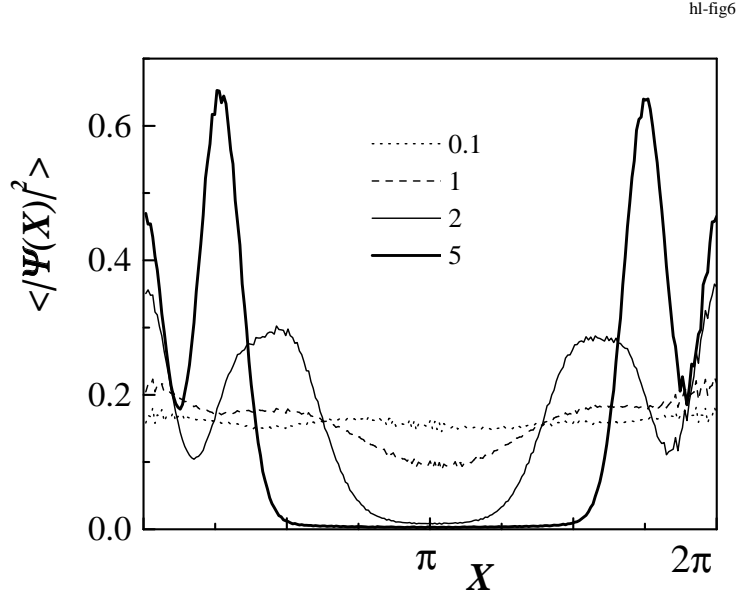,width=4.in}}
\vspace{-6cm}
\caption{
The wave function of an incommensurate ground state with
winding number $\omega_n=89/144$ at fixed $\tilde{\hbar}=0.2$ for
different
values of 
$K$. $\langle|\Psi|^2\rangle$ is the probability of finding
the particle at
$X$. The curves are for $K=0.1, 1, 2$, and $5$, respectively.
The wave function becomes localized in the lower part of the potential as
$K$ is increased to 2 and 5. 
}
\label{wvfig}
\end{figure}

Since the external potential is periodic with period $2\pi$, we can
fold the wave function to this period and then take the average over all
the
particles in the interval $[0,2\pi]$.  This quantity gives the probability
of finding the particle at a given position $X$. We plot the
averaged probability density $\langle|\Psi|^2\rangle$ in Fig.
\ref{wvfig} with 144 particles embedded in 89 potentials for a fixed
$\tilde{\hbar}$ ($=0.2$) but different values of $K$. Here
$\langle|\Psi|^2\rangle$ is normalized, i.e., $\int_0^{2\pi}dX
\langle|\Psi(X)|^2\rangle =1$. We observe that, in the small $K$ regime,
the probability of finding the particles at any point of the potential is
almost the same (see, for example, the curve for $K=0.1$). This is quite
similar to the
unpinned phase in the classical case. However, as the coupling constant
increases, the probability of finding the particle in the upper part of
the potential 
decreases whereas the probability of finding the particle in the lower
part of the potential
increases, as is noticeable from the appearance of peaks in the curves. As
the coupling constant increases, the
probability of finding the particles at the top is almost zero (see, for
example, the curves for $K=2$ and $5$).

\subsection{Indicators of the Aubry transition}

\subsubsection{Disorder Parameter}

In the classical FK model, Coppersmith and Fisher \cite{CF83} have
proposed a ``disorder parameter'' to describe the transition from the
pinned phase to the unpinned phase.  This parameter is defined as the
minimum
distance of a particle from the top of the potential, $D
=\min_{i,n}|X_i^e - 2\pi(n+\frac{1}{2})|$. If
$D \neq 0$, the particles are pinned; when $D=0$, they are 
unpinned. This ``disorder parameter'' also measures the
discontinuity (or width of the biggest gap) of the hull function.  One
might want to use the same function to describe the quantum
crossover. For instance, one may define a very similar quantum disorder
parameter $D_q=\min_{i,n}|\bar{X}_i-2\pi(n+\frac{1}{2})|$, where
$\bar{X}_i$ is
the expectation value of the $i$th particle's position. However, this
parameter $D_q$ could not capture the crossover of the ground state
wave function. The reason is that, in the quantum case, the particle can
tunnel from one side of the potential to the other. Thus the gap in the
classical hull function does not survive quantum fluctuations (see
Refs. \cite{BGS89} and \cite{HLZ98}). It turns out that for large $K$,
$D_q$ might still be close to (or equal to) zero. Therefore, a new
parameter is needed to describe the wave function crossover.
To this end, we define the probability of finding the particles at
the potential top as a quantum ``disorder parameter'',
\begin{equation}
P_t =\frac{1}{N}\sum_{n=0}^{N-1} \int 
|\Psi(X)|^2\delta\left[X-2\pi(n+\frac{1}{2})\right]dX. 
\label{Pt}
\end{equation}

\begin{figure}
\centerline{\psfig{file=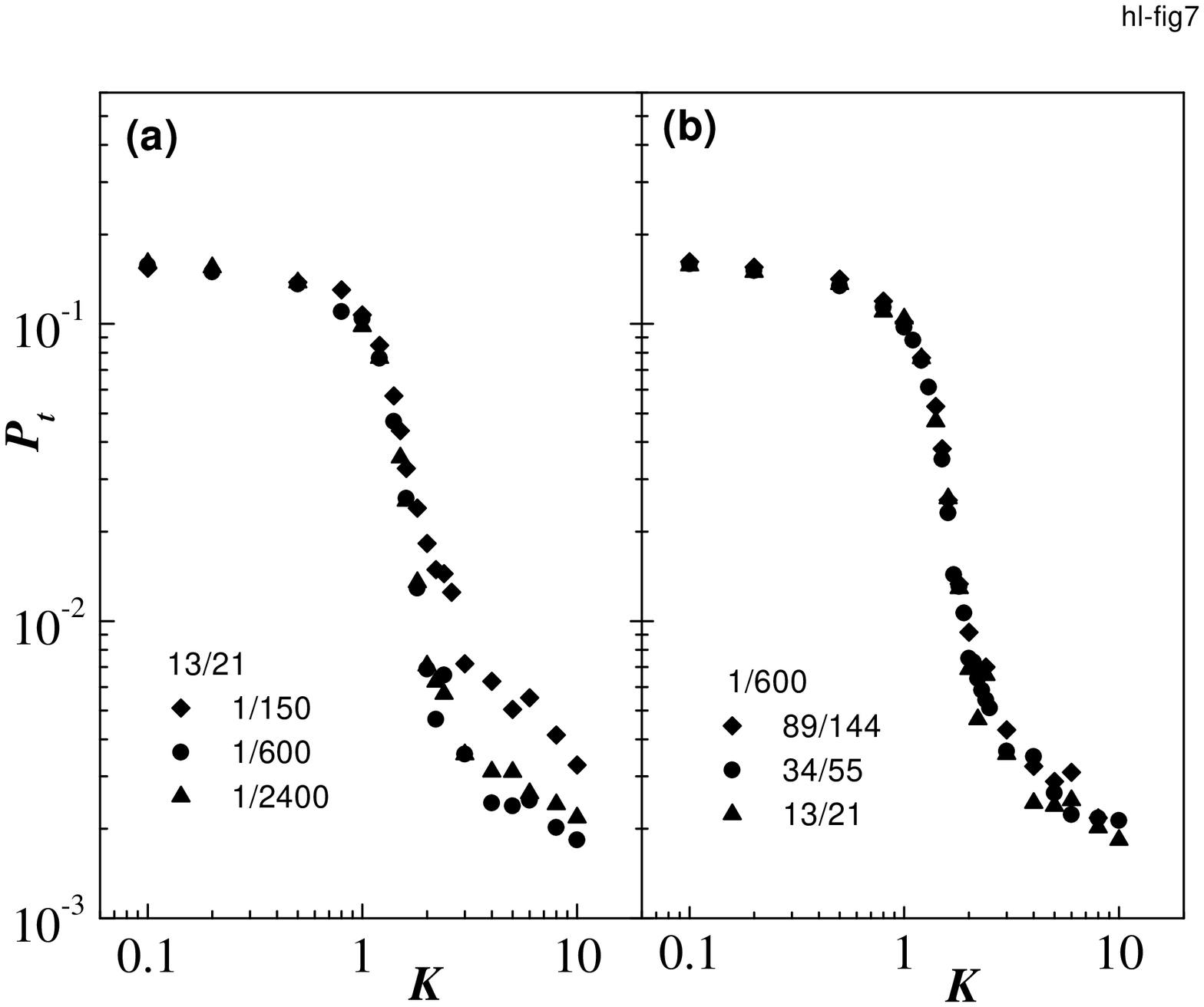,width=4.in}}
\vspace{-5.5cm}
\caption{The probability $P_t$
of finding the particles at the top of the potential as a function
of $K$ for
 $\tilde{\hbar}=0.2$.  (a) $P_t$ at
temperatures $T_e= 0.2/30, 0.2/120$, and $0.2/480$ for 
$\omega_n=13/21$. (b) $P_t$ for
$\omega_n = 13/21, 34/55$, and $89/144$ at a fixed temperature
$T_e=0.2/120$.
}
\label{qdisord}
\end{figure}

In Figs. \ref{qdisord}(a)-(b), we plot $P_t$ as a function of $K$ for
different
temperatures $T_e$ ($T_e=\tilde{\hbar}/\tau$, where $\tau$ is the
length of the imaginary time axis) and different system sizes. In Fig.
\ref{qdisord}(a), we fix $\tilde{\hbar}(=0.2)$, $\omega=13/21$ and set
the temperature 
$T_e=0.2/30, 0.2/120,$, $0.2/480$. The convergence is quite fast as is
seen from the figure. The sharp decrease of $P_t$ is very evident for 
$1< K< 2$. In this small $K$ regime, $P_t$ changes significantly: it
drops almost two orders of magnitude. This drastic change can also be
seen from Fig.  \ref{qdisord}(b). There the temperature is fixed at
$T_e=0.2/120$, which can be regarded as
effectively zero temperature.  The system size is changed from 21
particles to 144 particles.

\subsubsection{Maximum Variance}

\begin{figure} 
\centerline{\psfig{file=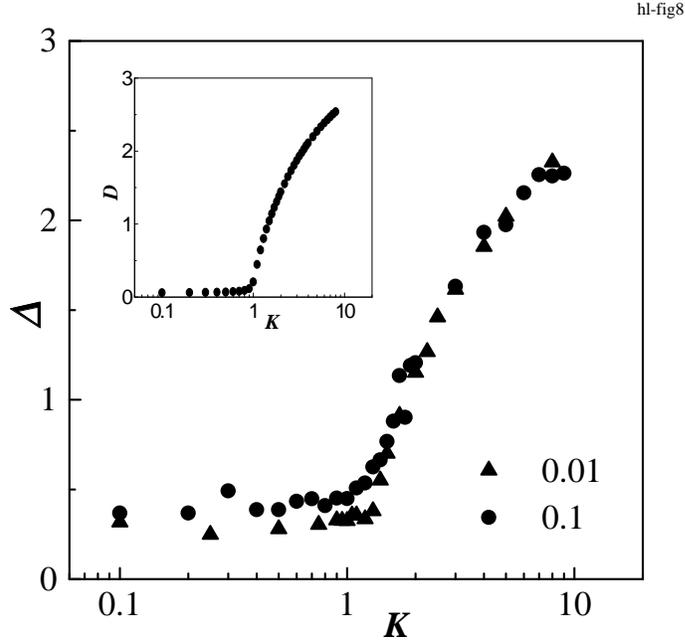,width=4.in}}
\vspace{-5.cm}
\caption{
The maximum variance $\Delta$ as a function of $K$ for $\omega_n=34/55$.
Different symbols
represent $\tilde{\hbar}= 0.01$ and $0.1$, respectively. We
draw the classical disorder parameter $D$
in the inset for comparison.
} 
\label{deltak}
\end{figure}

Another parameter that can be used to
depict this crossover is the maximal fluctuation of the particles. In the
quantum case, we have observed that the particle situated at a position
nearest to the top of the potential always has a maximal fluctuation since
it has a higher classical potential energy. This observation has been
verified numerically by QMC and theoretically
by the squeezed state theory \cite{HLZ98}.  Thus we can use this maximal
fluctuation as another measure of quantum ``disorder''. To make a
qualitative comparison with the classical disorder parameter $D$, we define
\begin{equation} 
\Delta = \max_i \left[\sqrt{\langle (X_i -
\bar{X}_i)^2\rangle}\right]. 
\label{DQM}
\end{equation} 
Here the average $\langle\cdots\rangle$ is taken over all
the paths (approximately 4,000) used in QMC.  $\Delta$ as a function 
of $K$ is
plotted in Fig.
\ref{deltak}. The computations given in this figure have been carried out
with $\omega_n=F_n/F_{n+1}=34/55$.

It is noticeable that the transition of the wave function is characterized
by the different $K$-dependent behavior of $\Delta$.  In the small $K$
regime,
$\Delta$ is a constant independent of $K$ but
changes with
$\tilde{\hbar}$. In the large $K$ regime, $\Delta$
increases with
$K$ but does not change with $\tilde{\hbar}$. Furthermore, $\Delta$
is approximately equal to the classical
disorder parameter $D$. For comparison, we have included $D$ in the
inset of Figure \ref{deltak}.
In the classical case, the transition of course occurs at $K=K_c$.

These results show that the width and the shape of the probability density
do not change with $K$ in the
small $K$ 
regime. The width only depends on the strength of the quantum
fluctuation $\tilde{\hbar}$. However, this picture changes significantly
for large $K$. In this regime, the profile of the
probability density
spreads out, and the width of the probability density is quite insensitive
to
quantum fluctuations and depends only on the coupling constant.  
It should however be stressed that the analogy between $D$ and
$\Delta$ cannot be
carried too far since $\Delta$ is not exactly the quantum correspondence
of $D$. This is why even if we let $\tilde{\hbar}$ go to zero,
$\Delta$ would not be zero in the small $K$ regime.

\subsubsection{Correlation function}

As another indication of the quantum signature of the Aubry transition,
we study the correlation function between the particles'
fluctuations.
The cross-correlation function $C_{ij}$  is defined by
\begin{equation}
C_{ij} = \frac{l_{ij}}{\sqrt{l_{ii} l_{jj}}} .
\label{CORR1}
\end{equation}
It describes the
fluctuation correlation between the particles $i$ and $j$.
$l_{ij} = \langle(X_i -\bar{X}_i)(X_j -\bar{X}_j)\rangle$
is the covariance of the fluctuations between the $i$th particle and the
$j$th particle. Here
$\langle\cdots\rangle$ denotes the ensemble average over space and `time'
in
QMC \cite{Linote}.
Since $l_{ij}$ is symmetric, $C_{ij}=C_{ji}$.
In our numerical calculations, we
take 4000 paths and each path is divided into many steps in time, ranging
from 1500 to 3000 depending on $\tilde{\hbar}$. The convergence is
guaranteed by increasing the number of steps and the number of paths until
the same results are reached.

\begin{figure}
\centerline{\psfig{file=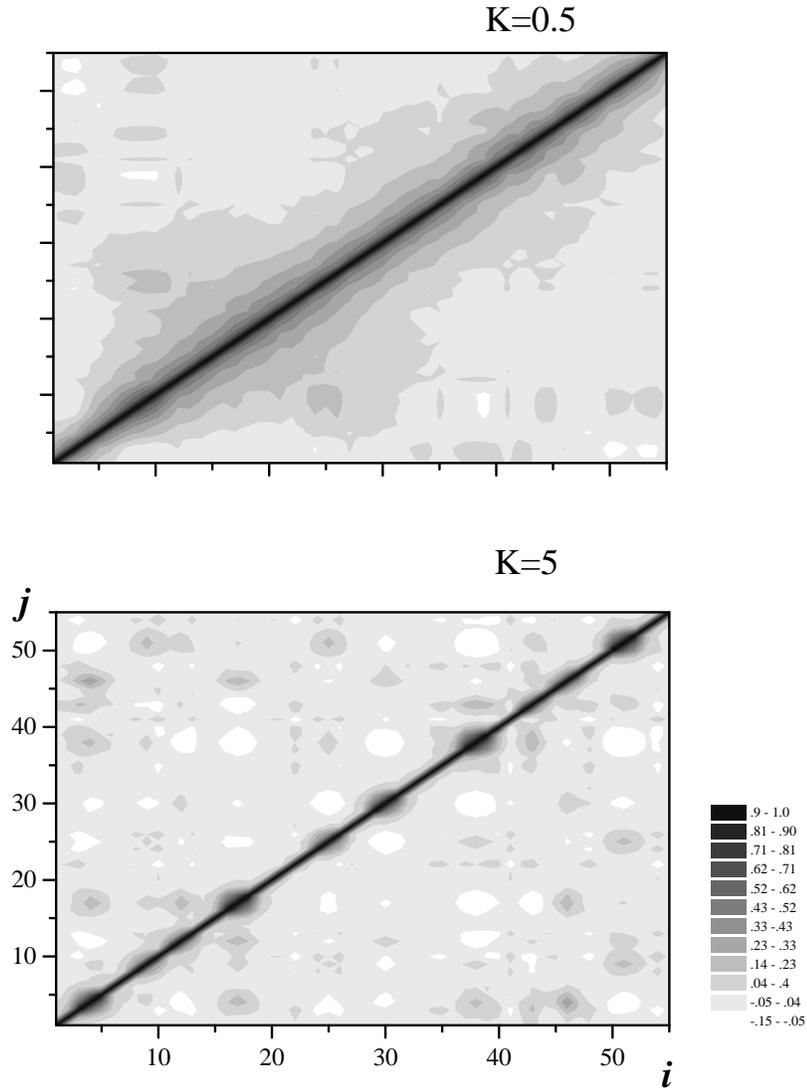,width=4.5in}}
\vspace{-1cm}
\caption{
Contour plot of the correlation function $C_{ij}$ for $K=0.5$ and $K=5$
at $\tilde{\hbar}=1$.
}
\label{corij}
\end{figure}

\begin{figure}
\centerline{\psfig{file=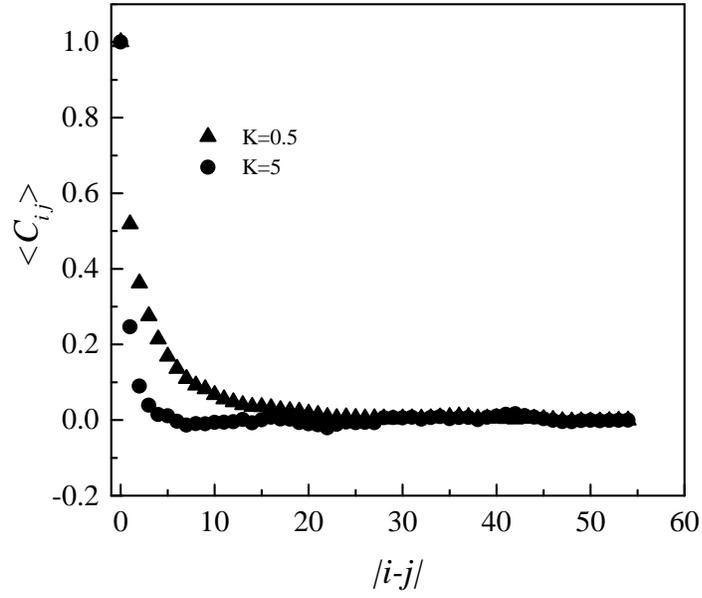,width=4.in}}
\vspace{-5.5cm}
\caption{
$\langle C_{ij}\rangle$ as a function of $|i-j|$. The average is taken
over the 
range of $i$ and $j$, $\tilde{\hbar}=1$.
}
\label{avgcor}
\end{figure}

Figure \ref{corij} shows the contour plots of $C_{ij}$ for 
$K=0.5$ and $K=5$. The plots are drawn in 12 equal steps from $-0.15$ to
$1$. A negative correlation function implies that the
particles' quantum fluctuations are out of phase.  The sharp change of
$C_{ij}$ from $K=0.5$ to $K=5$ is very readily noted. In the large $K$
regime, the correlation function matrix $C_{ij}$ is almost diagonal with
a very small band width whereas in the small $K$ regime the number
of nonzero off-diagonal elements increases and the band width is much
larger. Thus, in the small $K$ regime, we have a much longer range
interaction than in the large $K$ regime. This can also be seen
from the averaged correlation function $C(|i-j|)=\langle C_{ij}\rangle$,
where
$\langle\cdots\rangle$ denotes the ensemble average over particles.  We
have investigated this function for a wide range of $K$. We tried  to use
different forms to best-fit
this function, but
found that it is very hard to see whether the decay is exponential,
power-law or logarithmic. It seems that it obeys different laws in
different ranges. Nevertheless, the
correlation function in the large $K$ regime decays faster than that
in the small $K$ regime, as illustrated 
in Fig. \ref{avgcor} for $K=0.5$ and $K=5$, respectively. 

\begin{figure}
\centerline{\psfig{file=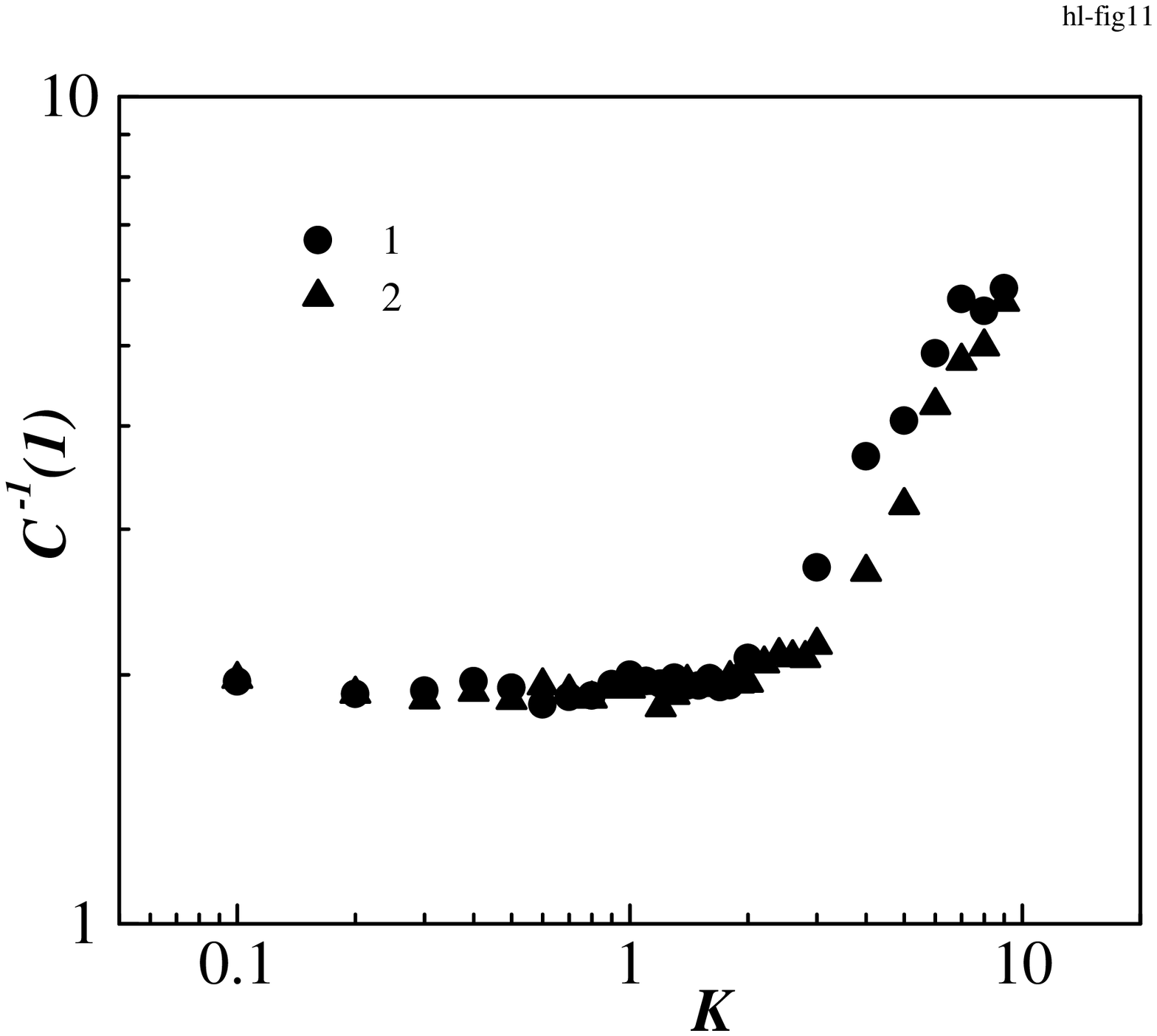,width=4.in}}
\vspace{-6.cm}
\caption{$C^{-1}(1)$ as a function of $K$
 for $\tilde{\hbar}=1$ and $2$, respectively.}
\label{c1}
\end{figure}

On the other hand, due to discreteness, it is also very difficult to   
define a parameter such as the width of the correlation function. As
an alternative, we chose $C(1)$, which describes the correlation
between the nearest-neighbor particles.
In Fig. \ref{c1}, we plot $C^{-1}(1)$ against $K$ on a logarithmic scale. 
The signature of the transition is very similar to Fig. \ref{deltak}. In
the small $K$ regime,
$C(1)$ is a constant (about 0.5); and in the large $K$ regime, 
it decreases with $K$. 
Since the transition  is largely smeared by quantum fluctuations, one
can't define a critical $K_c$ as in the classical case.

We know that if the correlation function decays exponentially in the short
range, then $\ln C^{-1}(1)= l^{-1}$, which is just the
Lyapunov exponent $\sigma$ in the classical case \cite{Aubry}, $l$ being
the
correlation
length. Thus, the parameter $C(1)$ captures the short-range correlation.
Fig. \ref{c1} mimics the classical plot of the correlation length 
proposed by Aubry \cite{Aubry}. In the 
classical case, for $K<K_c$, $l=\infty$ and
$\sigma=0$, whereas for $K>K_c$, $l$ is finite and
$\sigma>0$. However, 
in the quantum FK model, 
we do not have infinitely long interaction in the small $K$ regime.
Instead, we have a finite interaction range, but it is much
longer than that in the large $K$ regime.

\section{Concluding remarks}

The Frenkel-Kontorova model is a simple model. However, despite its
deceptively simple appearance, it exhibits very complex behavior and
contains very rich physics. It has found applications in a wide range of
fields, from nonlinear dynamics to condensed matter physics to
tribology.  However, most of the studies have so far been confined to the
classical Frenkel-Kontorova model. Very little work has been done on the
quantum Frenkel-Kontorova model. We think this is indeed quite
regrettable.
The quantum Frenkel-Kontorova model is not only interesting from a
theoretical point of view but is also important  from a practical  point
of view. The work reported in this review represents only a first step in
this direction. We however hope that it will serve as an impetus to
further
studies. It is our conviction that a thorough understanding  of the
quantum
Frenkel-Kontorova model will lead to important breakthroughs\ both in
theory
and applications.

\begin{ack}
We would like to thank S. Aubry, F. Borgonovi, D. K. Campbell, H. 
Chen, R. B.  Griffiths, C.-L. Ho, H.-Q. Lin, D. Shepelyansky, L.-H. Tang,
and W.-M.
Zhang for many
helpful discussions.
This work was supported
in part by  grants from the Hong Kong Research Grants Council (RGC) and
the Hong Kong Baptist University Faculty Research Grant (FRG). 
\end{ack}

\newpage


\begin{thebibliography}{99}

\bibitem{FK38}
Y.~Frenkel and T.~K.~Kontorova, {\em Zh.~Eksp.~Teor.~Fiz.} {\bf 8}
(1938) 1340; F.~Frank and J.~van der Merwe, {\em Proc. R. Soc. London}  
{\bf 198} (1949) 205.

\bibitem{Aubry} 
S. Aubry, in {\em Solitons and Condensed Matter Physics}, 
edited by A. R. Bishop and T. Schneider. Springer-Verlag, Berlin 1978; 
{\em J. Phys. (Paris)} {\bf 44} (1983) 147; M. Peyrard and S. Aubry, {\em J. 
Phys.} C {\bf 16} (1983) 1593; S. Aubry, {\em Physica} D {\bf 7} (1983) 
240; S. Aubry and P.Y. LeDaeron, {\it ibid} {\bf 8} (1983) 381.


\bibitem{tribology}
M. G. Rozman, M. Urbakh, and J. Klafter, {\em Phys. Rev. Lett.} {\bf 77} 
(1996) 683; {\em Phys. Rev.} E {\bf 54}, (1996) 6485; M. Weiss 
and F. J. Elmer, {\em Phys. Rev.} B {\bf 53} (1996) 7539; T. Gyalog and
H. 
Thomas, {\em Europhys. Lett.} {\bf 37} (1996) 195;
O.~M.~Braun, T.~Dauxois, M.~V.~Paliy, and M.~Peyrard, {\em Phys. Rev. Lett.}
{\bf 78} (1997) 1295, {\em Phys. Rev.} E {\bf 55} (1997) 3598;
O.~M.~Braun,
A.~R.~Bishop, and J.~R\"oder, {\em Phys. Rev. Lett.} {\bf 79} (1997) 3692.
T.~Stunz and F. J Elmer, {\em Phys. Rev. } E {\bf 58} (1998) 1602, 1612.

\bibitem{Klafter}M.~G.~Rozman, M.~Urbakh, and J.~Klafter, {\em Phys. Rev. 
Lett.} {\bf 77} (1996) 683; V.~Zaloj, M.~Urbakh, and J.~Klafter, {\em Phys. 
Rev. Lett.} {\bf 81} (1998) 1227.

\bibitem{BGS89} 
F. Borgonovi, I. Guarneri and D. Shepelyansky,
{\em Phys. Rev. Lett.} {\bf 63} (1989) 2010; {\it ibid.}  {\em Z. Phys.}  
B {\bf 79} (1990) 133,
and F. Borgonovi,  Ph.D dissertation, Universit\'a Degli Studi di 
Pavia, 1989, Italia.

\bibitem{BBC94} 
G. P. Berman, 
E. N. Bulgakov and D. K. Campbell, {\em Phys. Rev.} B {\bf 49}, (1994)
8212.

\bibitem{HLZ98}B. Hu, {\em Quantum and Classical Correspondence:
Proceedings of the 4th Drexel Symposium on Quantum Nonintegrability},
p.
445, eds. D. H. Feng and B. L. Hu. International Press (Cambridge, MA)
1997; 
B.~Hu, B.~Li, and W.-M.~Zhang, {\em Phys. Rev.} E {\bf 58} (1998) R4068.

\bibitem{HL99} B. Hu and B. Li, {\em Europhys. Lett.} {\bf 46} (1999) 655.

\bibitem{ZF905} 
W. M. Zhang, D. H. Feng and R. Gilmore, {\em Rev. Mod. Phys.} 
{\bf 62} (1990) 867; W. M. Zhang and D. H. Feng, {\em Phys. 
Rep.} {\bf 252} (1995) 1, and the references therein.


\bibitem{Tsue91} 
Y. Tsue and Y. Fujiwara, {\em Prog. Theor. Phys.} {\bf 86} (1991) 443,
469.

\bibitem{Tsue92}
Y.Tsue, {\em Prog. Theor. Phys.} {\bf 88} (1992) 911.

\bibitem{PS97}
A. K. Pattanayak and W. C. Schieve, {\em Phys. Rev. } E {\bf 56} (1997)
278.

\bibitem{HLLZ98}
B. Hu, B. Li, J. Liu, and J.-L Zhou, {\em Phys. Rev.} E {\bf 58} (1998)
1743.

\bibitem{Linote}
This is slightly different from 
that of Borgonovi {\it et al.} \cite{BGS89}. There they set $X_0 = 0, X_Q
=  2\pi P$. We have calculated by QMC for both cases  and found
that the 
numerical results are slightly 
different. We use the periodic boundary condition.


\bibitem{QMC}M. Creutz and B. Freedman, {\em Ann. Phys.} {\bf 132}    
(1981) 472; E. V. Shuryak and O. V. Zhirov, {\em Nucl. Phys.} B {\bf  242} 
(1984) 393.

\bibitem{RMP97}S.~L.~Sondhi, S.~M.~Girvin, J.~P.~Carini, and D.~Shahar, 
Rev. Mod. Phys. {\bf 69} (1997) 315.

\bibitem{CF83}S. N.~Coppersmith and D.~S.~Fisher {\em Phys. Rev.} B {\bf
28} 
(1983) 2566.

\end{thebibliography}
\end{document}